\documentclass[journal]{IEEEtran} 
\IEEEoverridecommandlockouts

\usepackage{mathrsfs}
\usepackage[noadjust]{cite}
\usepackage{graphicx,color,overpic,psfrag}
\usepackage{amsmath, amssymb}
\usepackage{latexsym}
\usepackage{bm}
\usepackage{amssymb}
\usepackage{cases}
\usepackage{array}
\usepackage{fancyhdr}
\usepackage{setspace}
\usepackage{subfigure}
\usepackage{caption}
\usepackage{indentfirst}
\usepackage{cases}
\usepackage{url}
\usepackage{algpseudocode}
\usepackage{algorithm}
\usepackage{blkarray}

\usepackage{booktabs}
\usepackage{multirow}
\usepackage{dsfont}
\usepackage{tabularx}
\usepackage[table]{xcolor}
\usepackage{amsfonts}
\usepackage{amsthm}
\usepackage{stfloats}
\usepackage{letltxmacro}

\newcommand{\bg}{\mathbf{g}}

\newcommand{\br}{\mathbf{r}}

\newcommand{\bt}{\mathbf{t}}

\newcommand{\bB}{\mathbf{B}}
\newcommand{\bC}{\mathbf{C}}

\newcommand{\bI}{\mathbf{I}}

\newcommand{\bM}{\mathbf{M}}
\newcommand{\bN}{\mathbf{N}}

\newcommand{\bP}{\mathbf{P}}
\newcommand{\bQ}{\mathbf{Q}}

\graphicspath{{figure/}}
\allowdisplaybreaks

\begin{document}
	
\title{\LARGE Fluid Antenna-Assisted MIMO Transmission Exploiting Statistical CSI}

\author{ Yuqi~Ye, Li~You,~\IEEEmembership{Senior~Member,~IEEE,} Jue Wang,~\IEEEmembership{Member,~IEEE,} Hao Xu,~\IEEEmembership{Member,~IEEE,}\\ Kai-Kit Wong,~\IEEEmembership{Fellow,~IEEE,} 
and Xiqi~Gao,~\IEEEmembership{Fellow,~IEEE}
\vspace{-5mm}		

\thanks{

Yuqi Ye, Li You, and Xiqi Gao are with the National Mobile Communications Research Laboratory, Southeast University, Nanjing 210096, China, and also with the Purple Mountain Laboratories, Nanjing 211100, China (e-mail:  yqye@seu.edu.cn; lyou@seu.edu.cn; xqgao@seu.edu.cn).

Jue Wang is with the School of Information Science and Technology, Nantong University, Nantong 226019, China, and also with the Nantong Research Institute for Advanced Communication Technologies, Nantong 226019, China (e-mail: wangjue@ntu.edu.cn).

Hao Xu and Kai-Kit Wong are with the Department of Electronic and Electrical Engineering, University College London, United Kingdom (e-mail: hao.xu@ucl.uk;  kai-kit.wong@ucl.ac.uk).
}
}

\maketitle

\begin{abstract}
In conventional multiple-input multiple-output (MIMO) communication systems, the positions of antennas are fixed. To take full advantage of spatial degrees of freedom, a new technology called fluid antenna (FA) is proposed to obtain higher achievable rate and diversity gain. Most existing works on FA exploit instantaneous channel state information (CSI). However, in FA-assisted systems, it is difficult to obtain instantaneous CSI since changes in the antenna position will lead to channel variation.
In this letter, we investigate a FA-assisted MIMO system using relatively slow-varying statistical CSI. Specifically, in the criterion of rate maximization, we propose an algorithmic framework for transmit precoding and transmit/receive FAs position designs with statistical CSI. Simulation results show that our proposed algorithm in FA-assisted systems significantly outperforms baselines in terms of rate performance.

\end{abstract}

\begin{IEEEkeywords}
Antenna position optimization, fluid antenna system, MIMO, statistical CSI.
\end{IEEEkeywords}

\section{Introduction}
Multiple-input multiple-output (MIMO) systems have been widely used as a core technology for wireless communications because of its ability to improve the reliability and capacity \cite{wang2023road}. The conventional MIMO systems typically use fixed-position antennas (FPAs). Recently, fluid antenna (FA) technology has been proposed to further improve the diversity and multiplexing performance gains \cite{wong2020fluid2,shojaeifard2022mimo,wong2023fluid00}. In a FA-assisted system, antennas can move\footnote{More precisely, FA should be understood as an antenna that can change its position and such change may not necessarily involve physical movement of an antenna. In practice, this may be more suitably achieved by switching on or off the units in an array of compact radio-frequency pixels.} freely within a given region instead of being fixed at specific positions. By utilizing more degrees of freedom in the spatial domain, FA and MIMO can combine to achieve a much higher spatial diversity gain \cite{new2023information}.


Several studies have emerged to investigate the performance of FA-assisted systems. In \cite{wong2020fluid2}, the authors proposed the new fluid antenna system (FAS) and studied the performance of the single-antenna FAS. In \cite{wong2020performance}, the authors indicated that a single-antenna FAS in a tiny space can reach the capacity of maximum ratio combining with multiple antennas. In \cite{mukherjee2022level}, the average level crossing rate in FAS was given as closed-form expressions. In \cite{ma2022mimo}, the authors proposed the movable antenna-assisted MIMO system, which is effectively a MIMO FAS, and studied the capacity maximization problem based on instantaneous channel state information (CSI). In \cite{zhu2022modeling}, the authors investigated the channel modeling and performance analysis for movable antenna-assisted MIMO systems. 
Note that most existing works on FA exploited instantaneous CSI, which is a strong assumption since it is usually difficult to acquire instantaneous CSI in FA-assisted systems as changes in the antenna position will cause channel variation \cite{skouroumounis2022fluid,ma2023compressed}. On the other hand, the slow-changing property of statistical CSI makes it relatively easy to obtain. Thus, for practical consideration, statistical CSI can be exploited to facilitate the transmit design of FA-assisted systems \cite{you2020energy,9238778,you2020spectral}.


In this letter, we investigate a FA-assisted MIMO system exploiting statistical CSI. Specifically, we study the rate maximization problem by jointly optimizing the transmit covariance, the receive FA position, and the transmit FA position. First, Jensen's inequality is adopted to simplify the objective function. Then, we apply the alternating optimization method and divide the transformed problem into three sub-problems. We provide a closed-form solution for the transmit covariance sub-problem, while a second-order Taylor expansion is used to find a sub-optimal solution for the transmit/receive FA position sub-problem. Finally, simulation results demonstrate that the FA-assisted system ensures superior performance over the conventional FPA-assisted system.

\textit{Notations}: 
$\mathbb{E}\{\cdot\}$ represents the expectation operation. 
The notation $\operatorname{tr}\{\cdot\}$ means the trace, and $\triangleq$ is used for definitions. Additionally, $\mathbf{A}^H$ represents conjugate transpose of matrix $\mathbf{A}$ while $\mathbf{a}^T$ represents transpose of vector $\mathbf{a}$ and $\mathbf{A}_{i, j}$ denotes the $(i,j)$-th element of matrix $\mathbf{A}$. $\mathcal{CN}(\bf 0,\bB)$ is complex circularly symmetric Gaussian distribution with zero mean and covariance matrix $\bB$. $||\cdot||_2$ represents the $l_2$ norm. 
	
	
\section{System Model And Problem Formulation}
We consider a MIMO system that consists of a transmitter with $N$ FAs and a receiver with $M$ FAs. FAs at the transmitter and receiver are all connected to radio frequency chains via flexible cables so that they can move freely in the given regions $\mathcal {S}_{\rm t}$ and $\mathcal S_{\rm r}$, respectively. To represent the precise positions of the transmit and receive FAs, we introduce the two-dimensional Cartesian coordinate system. The coordinates of the $n$-th $(1\leq n\leq N)$ transmit FA and the $m$-th $(1\leq m\leq M)$ receive FA are denoted as $\mathbf t_n=(x_{n},y_{n})^T\in \mathcal S_{\rm t}$ and $\mathbf  r_m=(x_{m},y_{m})^T\in \mathcal S_{\rm r}$, respectively. We assume that both $\mathcal S_{\rm t}$ and $\mathcal S_{\rm r}$ are $A\times A$ square regions, i.e., $\mathcal S_{\rm t}=\mathcal S_{\rm r}=[-A/2,A/2]\times[-A/2,A/2]$. The position collections of $N$ transmit FAs and $M$ receive FAs are denoted by ${\mathbf  t}=[\mathbf  t_1,\dots,\mathbf  t_N]\in\mathbb R^{2\times N}$ and ${\mathbf  r}=[\mathbf  r_1,\dots,\mathbf  r_M]\in\mathbb R^{2\times M}$, respectively.


The transmit signal is denoted by $\mathbf s\in\mathbb C^{N\times 1}$. The transmit covariance matrix is defined by $\mathbf Q\triangleq\mathbb E\{\mathbf s \mathbf s^H\}\in\mathbb C^{N\times N}$. The receive signal is expressed as
\begin{equation}
\mathbf y({\mathbf t},{\mathbf r})=\mathbf H({\mathbf t},{\mathbf r})\mathbf s+\mathbf z,
\end{equation}
where $\mathbf H({\mathbf t},{\mathbf r})\in\mathbb C^{M\times N}$ is the channel matrix between the transmitter with $N$ FAs at positions $\bt$ and receiver with $M$ FAs at positions $\br$, and $\mathbf z\in\mathbb C^{M\times 1}\sim\mathcal{CN}(0,\sigma^2\mathbf I_M)$ is the complex additive white Gaussian noise.

We assume that the size of the FAs moving region is much smaller than the distance between the transmitter and the receiver so that the far-field model can be assumed \cite{ma2022mimo}. In this case, the channel is modeled assuming planar wavefront, which ignores the amplitude difference between the received signals of each array element, and considers that the received signals are simple time-delay relations. Moreover, for each channel path, the angles of departure (AoDs) and arrival (AoAs) depend mainly on the scatterer and propagation environment and do not vary with different antenna positions.
		
The number of transmit and receive paths is denoted by $L_{\rm_t}$ and $L_{\rm_r}$, respectively. For the transmitter side, the elevation and azimuth AoDs of the $p$-th $(1\leq p\leq L_{\rm_t})$ transmit path are, respectively, denoted as $\theta_{\rm t}^p\in[0,\pi]$ and $\phi_{\rm t}^p\in[0,\pi]$. In the $p$-th transmit path, the propagation distance difference between the position of the $n$-th transmit FA and origin $\mathbf t_0=(0,0)^T$ is \cite{ma2022mimo}
\begin{equation}
\rho_{\rm t}^p(\mathbf{t}_n)=x_n \sin \theta_{\rm t}^p \cos \phi_{\rm t}^p+y_n \cos \theta_{\rm t}^p.
\end{equation}
Correspondingly, the signal phase difference between the position of the $n$-th transmit FA and origin $\mathbf t_0=(0,0)^T$ in the $p$-th transmit path can be obtained as $2\pi\rho_{\rm t}^p(\mathbf{t}_n)/\lambda$, where $\lambda$ is the signal wavelength. Thus, the transmit field response vector can be written as
\begin{equation}
\mathbf{g}(\mathbf t) \triangleq\left[e^{\jmath\frac{2\pi}{\lambda}\rho_{\rm t}^1(\mathbf{t})},\dots,e^{\jmath\frac{2\pi}{\lambda}\rho_{\rm t}^{L_{\rm_t}}(\mathbf{t})}\right]^T \in \mathbb{C}^{L_{\rm_t}\times 1}.
\end{equation}
The field response matrix of all $N$ transmit FAs can be denoted as
\begin{equation}
\mathbf{G}({\mathbf{t}}) \triangleq\left[\mathbf{g}\left(\mathbf{t}_1\right), \mathbf{g}\left(\mathbf{t}_2\right), \ldots, \mathbf{g}\left(\mathbf{t}_N\right)\right] \in \mathbb{C}^{L_{\rm_t} \times N}.
\end{equation}

For the receiver side, the elevation and azimuth AoAs of the $q$-th $(1\leq q\leq L_{\rm_r})$ receive path are, respectively, denoted as $\theta_{\rm r}^q\in[0,\pi]$ and $\phi_{\rm r}^q\in[0,\pi]$. In the $q$-th receive path, the propagation distance difference between the position of the $m$-th receive FA and origin $\mathbf r_0=(0,0)^T$ is
\begin{equation}
\rho_{\rm r}^q(\mathbf{r}_m)=x_m \sin \theta_{\rm r}^q \cos \phi_{\rm r}^q+y_m \cos \theta_{\rm r}^q,
\end{equation}
and the corresponding path phase difference can be obtained as $2\pi\rho_{\rm r}^q(\mathbf{r}_m)/\lambda$. Thus, the receive field response vector can be written as
\begin{equation}
\mathbf{f}(\mathbf r) \triangleq\left[e^{\jmath\frac{2\pi}{\lambda}\rho_{\rm r}^1(\mathbf{r})},\dots,e^{\jmath\frac{2\pi}{\lambda}\rho_{\rm r}^{L_{\rm_r}}(\mathbf{r})}\right]^T \in \mathbb{C}^{L_{\rm_r}\times 1}.
\end{equation}
The field response matrix of all $M$ receive FAs can be denoted as
\begin{equation}
\mathbf{F}({\mathbf{r}}) \triangleq\left[\mathbf{f}\left(\mathbf{r}_1\right), \mathbf{f}\left(\mathbf{r}_2\right), \ldots, \mathbf{f}\left(\mathbf{r}_M\right)\right] \in \mathbb{C}^{L_{\rm_r} \times M}.
\end{equation}

The path response matrix from the origin of the transmit region $\mathbf t_0=(0,0)^T$ to the origin of the receive region $\mathbf r_0=(0,0)^T$ is defined as $\boldsymbol\Sigma\in \mathbb{C}^{L_{\rm_r} \times L_{\rm_t} }$. $\boldsymbol\Sigma_{q,p}$ refers to the response coefficient between the $p$-th transmit path and the $q$-th receive path. We assume $\boldsymbol\Sigma_{q,p}$ is independently and identically distributed, which is modeled as a Gaussian distributed random variable with zero mean and variance $\alpha^2$. Therefore, the channel matrix $\mathbf H({\mathbf t},{\mathbf r})$ from transmitter with FAs at positions $\bt$ to receiver with FAs at positions $\br$ can be written as \cite{ma2022mimo,8926431}
\begin{equation}
\mathbf H({\mathbf t},{\mathbf r})=\mathbf{F}^H({\mathbf{r}})\boldsymbol\Sigma\mathbf{G}({\mathbf{t}}).
\end{equation}
By exploiting the property that AoAs/AoDs usually vary much more slowly than path gains over time-varying channels \cite{9505311,8409338}, the ergodic achievable rate under transmit covariance $\bQ$, transmit FAs positions $\bt$, and receive FAs positions $\br$ is written as
\begin{equation}
\begin{aligned}
R=\mathbb{E}_{\boldsymbol\Sigma}\left\{\log \operatorname{det}\left(\mathbf{I}_M+\frac{1}{\sigma^2} \mathbf{H}({\mathbf{t}}, {\mathbf{r}}) \mathbf{Q} \mathbf{H}^H({\mathbf{t}}, {\mathbf{r}})\right) \right\}.\label{rate_ex}
\end{aligned}
\end{equation}

In this letter, we aim to maximize the achievable rate described in \eqref{rate_ex} using statistical CSI. The optimization problem is formulated as
\begin{equation}
\begin{aligned}
\max _{ \mathbf{Q}, {\mathbf{t}}, {\mathbf{r}}} 
&\quad \mathbb{E}_{\boldsymbol\Sigma}\left\{\log \operatorname{det}\left(\mathbf{I}_M+\frac{1}{\sigma^2} \mathbf{H}({\mathbf{t}}, {\mathbf{r}}) \mathbf{Q} \mathbf{H}^H({\mathbf{t}}, {\mathbf{r}})\right) \right\}\\
\text { s.t. } &\quad {\mathbf{t}} \in \mathcal{S}_{\rm t}, \\
& \quad {\mathbf{r}} \in \mathcal{S}_{\rm r}, \\
&\quad \left\|\mathbf{t}_k-\mathbf{t}_l\right\|_2 \geq D, \quad k, l=1,2, \ldots, N, \quad k \neq l, \\
&\quad\left\|\mathbf{r}_k-\mathbf{r}_l\right\|_2 \geq D, \quad k, l=1,2, \ldots, M, \quad k \neq l, \\
&\quad\operatorname{tr}(\mathbf{Q}) \leq P_{\text{max}},\label{ori_p}
\end{aligned}
\end{equation}
where $D$ is defined as the minimum required distance between transmit/receive FAs to avoid mutual coupling. 

Note that the optimization problem in \eqref{ori_p} involves a non-concave objective function with expectation operation, non-concave minimum required distance constraints, and coupled variables. These challenges make it difficult to tackle.

\section{Statistial CSI-aided Rate Maximization}
Here, we first use Jensen's inequality to derive an analytical upper bound for \eqref{rate_ex}. Then we propose an algorithm to convert the nonconvex problem \eqref{ori_p} into several convex sub-problems. Finally, we discuss the convergence and computational complexity of the proposed rate maximization algorithm. 

\subsection{Upper Bound of the Achievable Rate }
For \eqref{ori_p}, computing the expectation values is quite resource-consuming. Utilizing the traditional Monte Carlo method to manipulate the expectation operation is also computationally cumbersome. Therefore, we replace the original objective function with its tight upper bound. Jensen's inequality \cite{mckay2006random} can be exploited to derive the upper bound of $R$, given by 
\begin{equation}
\begin{aligned}
&R\leq \overline{R}
\triangleq\log \operatorname{det}\left(\mathbf{I}_M+\frac{1}{\sigma^2} \mathbb{E}_{\boldsymbol\Sigma}\left\{\mathbf{H}({\mathbf{t}}, {\mathbf{r}}) \mathbf{Q} \mathbf{H}^H({\mathbf{t}}, {\mathbf{r}}) \right\}\right)\\
&=\log\operatorname{det}\left(\mathbf{I}_M+\frac{1}{\sigma^2} \mathbf{F}^H({\mathbf{r}})\mathbb{E}_{\boldsymbol\Sigma}\left\{\boldsymbol\Sigma\bP\boldsymbol\Sigma^H\right\} \mathbf{F}({\mathbf{r}})\right),\label{upper}
\end{aligned}
\end{equation}
where we define $\bP\triangleq\mathbf{G}({\mathbf{t}})\mathbf{Q} \mathbf{G}^H({\mathbf{t}})$. Utilizing the statistical characteristics of $\boldsymbol\Sigma$, the expectation term in \eqref{upper} can be further written as \cite{mckay2006random}
\begin{equation}
\begin{aligned}
\mathbb{E}_{\boldsymbol\Sigma}\left\{\boldsymbol\Sigma\bP\boldsymbol\Sigma^H\right\}
=\operatorname{tr}(\bP)\alpha^2\bI_{L_{\rm_r}}.
\end{aligned}
\end{equation}
Correspondingly, the objective function can be rewritten as
\begin{equation}
\begin{aligned}
\overline{R}&=\log \operatorname{det}\left(\mathbf{I}_M+\frac{\alpha^2}{\sigma^2} \operatorname{tr}\left(\mathbf{G}({\mathbf{t}})\mathbf{Q} \mathbf{G}^H({\mathbf{t}})\right)\mathbf{F}^H({\mathbf{r}}) \mathbf{F}({\mathbf{r}})\right).\label{final_obj}
\end{aligned}
\end{equation}

Although the expectation operation is replaced, the problem is still very challenging since the optimization problem has a non-concave objective function, non-concave minimum required distance constraints and coupled variables. In the following, we exploit the alternating optimization method to handle the rate maximization problem. The idea of alternating optimization is to optimize the objective
function with respect to one variable while keeping the remaining variables fixed \cite{bertsekas1997nonlinear}. All variables are iterated until the
convergence condition is satisfied. The alternating optimization algorithm can at least find the local optimal solution of the original problem \cite{bertsekas1997nonlinear}.


\subsection{Optimization of Transmit Covariance Matrix }
With given $\mathbf r$ and $\mathbf t$, the transmit covariance matrix optimization sub-problem can be written as
\begin{equation}
\begin{aligned}
\max _{\mathbf{Q}} 
&\quad \log \operatorname{det}\left(\mathbf{I}_M+\frac{\alpha^2}{\sigma^2} \operatorname{tr}\left(\mathbf{G}({\mathbf{t}})\mathbf{Q} \mathbf{G}^H({\mathbf{t}})\right)\mathbf{F}^H({\mathbf{r}}) \mathbf{F}({\mathbf{r}})\right)\\
\text { s.t. } 
& \quad\operatorname{tr}(\mathbf{Q}) \leq P_{\text{max}}.\label{Q_op} 
\end{aligned}
\end{equation}

Because the derivative of $\log\det(\mathbf I+x\mathbf W)$ with respect to $x$ is always greater than 0, maximizing the objective function in \eqref{Q_op} is equivalent to maximizing $\operatorname{tr}\left(\mathbf{G}({\mathbf{t}})\mathbf{Q} \mathbf{G}^H({\mathbf{t}})\right)$. Thus, the sub-problem can be reformulated as
\begin{equation}
\begin{aligned}
\max _{\mathbf{Q}} 
&\quad \operatorname{tr}\left(\mathbf{G}({\mathbf{t}})\mathbf{Q} \mathbf{G}^H({\mathbf{t}})\right)\\
\text { s.t. } 
& \quad\operatorname{tr}(\mathbf{Q}) \leq P_{\text{max}}. \label{Q_sub}
\end{aligned}
\end{equation}

Since the trace operation obeys the Cauchy-Schwarz inequality, we have $\operatorname{tr}(\mathbf{M}\bN)\leq\sqrt{\operatorname{tr}(\mathbf{M}^2)\operatorname{tr}(\mathbf{N}^2)}$ \cite{yang2001matrix}. The equality holds when $\bM$ is a multiple of $\bN$. Applying this property to the objective of \eqref{Q_sub}, we have
\begin{equation}
\begin{aligned}
\operatorname{tr}\left(\mathbf{G}({\mathbf{t}})\mathbf{Q} \mathbf{G}^H({\mathbf{t}})\right)
&=\operatorname{tr}\left(\mathbf{Q}\mathbf{G}^H({\mathbf{t}}) \mathbf{G}({\mathbf{t}})\right)\\
&\leq\sqrt{\operatorname{tr}(\mathbf{Q}^2)\operatorname{tr}([\mathbf{G}^H ({\mathbf{t}})\mathbf{G}({\mathbf{t}})]^2)}.\label{CW}
\end{aligned}
\end{equation} 
When the objective function reaches its maximum value, it satisfies that $\bQ$ is a multiple of $\mathbf{G}^H ({\mathbf{t}})\mathbf{G}({\mathbf{t}})$. Moreover, the optimal solution of $\mathbf Q$ should also satisfy $\operatorname{tr}(\mathbf{Q}) =P_{\text{max}}$. 

\subsection{Optimization of Receive FA Position }
With given $\mathbf Q$ and $\mathbf t$, the receive FA position optimization sub-problem can be written as
\begin{equation}
\begin{aligned}
\max _{{\mathbf{r}}} 
&\quad \log\operatorname{det}\left(\mathbf{I}_M+\frac{\alpha^2}{\sigma^2} \operatorname{tr}\left(\mathbf{G}({\mathbf{t}})\mathbf{Q} \mathbf{G}^H({\mathbf{t}})\right)\mathbf{F}^H({\mathbf{r}}) \mathbf{F}({\mathbf{r}})\right)\\
\text { s.t. } 
& \quad{\mathbf{r}} \in \mathcal{S}_{\rm r}, \\
&\quad\left\|\mathbf{r}_k-\mathbf{r}_l\right\|_2 \geq D, \quad k, l=1,2, \ldots, M, \quad k \neq l. 
\end{aligned}
\end{equation}

Let $a=\frac{\alpha^2}{\sigma^2} \operatorname{tr}\left(\mathbf{G}({\mathbf{t}})\mathbf{Q} \mathbf{G}^H({\mathbf{t}})\right)$. The objective function $\overline{R}$ can be rewritten as
\begin{equation}
\begin{aligned}
\overline{R}
&=\log \operatorname{det}\left(\mathbf{I}_M+a\mathbf{F}^H({\mathbf{r}}) \mathbf{F}({\mathbf{r}})\right)\\
&=\log \operatorname{det}\left(\mathbf{I}_{L_{\rm_r}}+a\mathbf{F}({\mathbf{r}}) \mathbf{F}^H({\mathbf{r}})\right)\\
&=\log \operatorname{det}\left(\mathbf{I}_{L_{\rm_r}}+a\sum_{m=1}^M\mathbf{f}({\mathbf{r}_m}) \mathbf{f}^H({\mathbf{r}_m})\right).\label{R_trans}
\end{aligned}
\end{equation}
The final expression in \eqref{R_trans} decouples the position variables of $M$ receive FAs, making it easier to optimize each FA position separately.

After removing the $m$-th column vector $\mathbf{f}({\mathbf{r}_m})$ from matrix $\mathbf{F}({\mathbf{r}})$, we denote the remaining $L_{\rm_r} \times (M-1)$ matrix as $\bar{\mathbf{F}}_ m=[\mathbf{f}\left(\mathbf{r}_1\right), \mathbf{f}\left(\mathbf{r}_2\right), \ldots, \mathbf{f}\left(\mathbf{r}_{m-1}\right), \mathbf{f}\left(\mathbf{r}_{m+1}\right),\dots, \mathbf{f}\left(\mathbf{r}_{M}\right)]$. After separating $\mathbf{F}({\mathbf{r}})$ into $\mathbf{f}({\mathbf{r}_m})$ and $\bar{\mathbf{F}}_ m$, the objective function $\overline{R}$ can be rewritten as \cite{ma2022mimo}
\begin{equation}
\begin{aligned}
\overline{R}&=\log\operatorname{det}\left(\mathbf{I}_{L_{\rm_r}}+a\left(\bar{\mathbf{F}}_m \bar{\mathbf{F}}_m^H+\mathbf{f}({\mathbf{r}_m})\mathbf{f}^H({\mathbf{r}_m})\right)\right)\\
&=\log \operatorname{det}\left(\mathbf{I}_{L_{\rm_r}}+a\left(\mathbf{I}_{L_{\rm_r}}+a\bar{\mathbf{F}}_m \bar{\mathbf{F}}_m^H\right)^{-1}\mathbf{f}({\mathbf{r}_m})\mathbf{f}^H({\mathbf{r}_m})\right)+\\
&\quad \log\operatorname{det}\left(\mathbf{I}_{L_{\rm_r}}+a\bar{\mathbf{F}}_m \bar{\mathbf{F}}_m^H\right)\\
&=\log \operatorname{det}\left(1+a\mathbf{f}^H({\mathbf{r}_m})\left(\mathbf{I}_{L_{\rm_r}}+a\bar{\mathbf{F}}_m \bar{\mathbf{F}}_m^H\right)^{-1}\mathbf{f}({\mathbf{r}_m})\right)+\\
&\quad\log\operatorname{det}\left(\mathbf{I}_{L_{\rm_r}}+a\bar{\mathbf{F}}_m \bar{\mathbf{F}}_m^H\right).
\end{aligned}
\end{equation}

With given $\{\mathbf{r}_k,k\neq m\}_{k=1}^M$, maximizing $\overline{R}$ equals maximizing $p({\mathbf{r}_m})=\mathbf{f}^H({\mathbf{r}_m})\bB_m\mathbf{f}({\mathbf{r}_m})$, where 
\begin{equation}
\begin{aligned}
\bB_m=\left(\mathbf{I}_{L_{\rm_r}}+a\bar{\mathbf{F}}_m \bar{\mathbf{F}}_m^H\right)^{-1}\label{Am}
\end{aligned}
\end{equation}
is a positive definite matrix independent of $\mathbf{r}_m$. Then the receive FA position optimization sub-problem can be reformulated as
\begin{equation}
\begin{aligned}
\max _{\mathbf{r}_m} 
&\quad p({\mathbf{r}_m})=\mathbf{f}^H({\mathbf{r}_m})\bB_m\mathbf{f}({\mathbf{r}_m})\\
\text { s.t. } 
&\quad {\mathbf{r}_m} \in \mathcal{S}_{\rm r}, \\
&\quad \left\|\mathbf{r}_m-\mathbf{r}_k\right\|_2 \geq D, \quad k=1,2, \ldots, M, \quad k \neq m, \label{r_sub}
\end{aligned}
\end{equation}
which can be handled by the second-order Taylor expansion approach described in \cite{ma2022mimo}.

\subsection{Optimization of Transmit FA Position }
With given $\mathbf Q$ and $\mathbf r$, the transmit FA position optimization sub-problem can be written as
\begin{equation}
\begin{aligned}
\max _{\mathbf{t}} 
&\quad \log\operatorname{det}\left(\mathbf{I}_M+\frac{\alpha^2}{\sigma^2} \operatorname{tr}\left(\mathbf{G}({\mathbf{t}})\mathbf{Q} \mathbf{G}^H({\mathbf{t}})\right)\mathbf{F}^H({\mathbf{r}}) \mathbf{F}({\mathbf{r}})\right)\\
\text { s.t. } &\quad {\mathbf{t}} \in \mathcal{S}_{\rm t}, \\
&\quad\left\|\mathbf{t}_k-\mathbf{t}_l\right\|_2 \geq D, \quad k, l=1,2, \ldots, N, \quad k \neq l. \label{t_ori}
\end{aligned}
\end{equation}

With $\mathbf{t}$ being the optimization variable, maximizing the objective function in \eqref{t_ori} is equivalent to maximizing $\operatorname{tr}\left(\mathbf{G}({\mathbf{t}})\mathbf{Q} \mathbf{G}^H({\mathbf{t}})\right)$. Thus, the sub-problem can be reformulated as
\begin{equation}
\begin{aligned}
\max _{\mathbf{t}} 
&\quad \operatorname{tr}\left(\mathbf{G}({\mathbf{t}})\mathbf{Q} \mathbf{G}^H({\mathbf{t}})\right)\\
\text { s.t. } &\quad {\mathbf{t}} \in \mathcal{S}_{\rm t}, \\
&\quad\left\|\mathbf{t}_k-\mathbf{t}_l\right\|_2 \geq D, \quad k, l=1,2, \ldots, N, \quad k \neq l. \label{t_op}
\end{aligned}
\end{equation}

Similarly, using the Cauchy-Schwarz inequality, the objective function in \eqref{t_op} is upper bounded by \eqref{CW}. Thus, maximizing the objective function in \eqref{t_op} is same as maximizing
\begin{equation}
\begin{aligned}
&\operatorname{tr}\left(\mathbf{G}^H({\mathbf{t}}) \mathbf{G}({\mathbf{t}})\mathbf{G}^H({\mathbf{t}}) \mathbf{G}({\mathbf{t}})\right)=\operatorname{tr}\left(\mathbf{G}({\mathbf{t}}) \mathbf{G}^H({\mathbf{t}})\mathbf{G}({\mathbf{t}}) \mathbf{G}^H({\mathbf{t}})\right)\\
&=\operatorname{tr}\left(\sum_{n=1}^N\bg(\bt_n)\bg^H(\bt_n)\sum_{m=1}^N\bg(\bt_m)\bg^H(\bt_m)\right).\label{tr_final}
\end{aligned}
\end{equation}

With given $\{\bt_k,k\neq n\}_{k=1}^N$, maximizing \eqref{tr_final} equals to maximizing $\operatorname{tr}\left(\bg(\bt_n)\bg^H(\bt_n)\bC_n\right)=\bg^H(\bt_n)\bC_n\bg(\bt_n)$, where 
\begin{equation}
\begin{aligned}
\bC_n=\sum_{m\neq n}^{N}\bg(\bt_m)\bg^H(\bt_m)\label{Bn}
\end{aligned}
\end{equation}
is a positive semidefinite matrix independent of $\bt_n$.
Then the transmit FA position optimization sub-problem can be reformulated as
\begin{equation}
\begin{aligned}
\max _{\mathbf{t}_n} 
&\quad q({\mathbf{t}_n})=\mathbf{g}^H({\mathbf{t}_n})\bC_n\mathbf{g}({\mathbf{t}_n})\\
\text { s.t. } 
&\quad {\mathbf{t}_n} \in \mathcal{S}_{\rm t}, \\
&\quad \left\|\mathbf{t}_n-\mathbf{t}_k\right\|_2 \geq D, \quad k=1,2, \ldots, N, \quad k \neq n, \label{t_sub}
\end{aligned}
\end{equation}
which can be handled by the second-order Taylor expansion described in \cite{ma2022mimo}.

\begin{algorithm}[t]
\caption{Rate Maximization Algorithm with Statistical CSI}
\label{a_1}
 \begin{algorithmic}[1]
\Require $M$, $N$, $L_{\rm_t}$, $L_{\rm_r}$, $\{\theta_{\rm t}^p\}_{p=1}^{L_{\rm_t}}$, $\{\phi_{\rm t}^p\}_{p=1}^{L_{\rm_t}}$, $\{\theta_{\rm r}^q\}_{q=1}^{L_{\rm_r}}$, $\{\phi_{\rm r}^q\}_{q=1}^{L_{\rm_r}}$, $\sigma^2$, $\alpha^2$, $\boldsymbol\Sigma$, $D$, $P_{\text{max}}$.
\Ensure $\bt$, $\br$, $\bQ$.
 \State Initialize $\bt$, $\br$, threshold $\varepsilon$, set iteration $i=0$, and calculate objective function $\overline R^{(0)}$ as \eqref{final_obj}.
\Repeat 
\State Update $\bQ$ by optimizing sub-problem \eqref{Q_sub}.
\For {m = 1: M}
\State Calculate $\bB_m$ as \eqref{Am}.
\State Update $\br_m$ by optimizing sub-problem \eqref{r_sub}.
\EndFor
\For {n = 1: N}
\State Calculate $\bC_n$ as \eqref{Bn}.
\State Update $\bt_n$ by optimizing sub-problem \eqref{t_sub}.
\EndFor
 \State Set $i=i+1$.
\State Calculate $\overline R^{(i)}$ as \eqref{final_obj}.
 \Until{$|\overline R^{(i)}-\overline R^{(i-1)}|\leqslant \varepsilon$.}\label{a1}
 \end{algorithmic}
\end{algorithm}

To summarize, a statistical CSI-aided rate maximization algorithm for FA-assisted systems is detailed in \textbf{Algorithm 1}. The proposed algorithm consists of outer and inner iterations. Concerning the inner iteration, the optimization sub-problems satisfy the convergence conditions mentioned in \cite{ma2022mimo}. For outer iterations, the objective function is bounded and non-decreasing. The proposed iterative algorithm can therefore be guaranteed to converge.

The computational complexity of \textbf{Algorithm 1} is analyzed as follows. In Step 3, the computational complexity of updating $\bQ$ is $\mathcal{O}\left(N^2 L_{\rm_t} \right)$. From Step 4 to Step 7, the corresponding complexity to compute receive FA position $\br$ is estimated as $\mathcal{O}\left(M N L_{\rm_r} \gamma_{\rm_r}^1+M^{2.5} \ln (1 / \beta) \gamma_{\rm_r}^2\right)$, where $\gamma_{\rm_r}^1$ and $\gamma_{\rm_r}^2$ are the maximum number of inner iterations and the maximum number of iterations required to handle the quadratic programming problem, respectively, and $\beta$ is the accuracy of the interior-point method \cite{ma2022mimo}. Similarly, the computational complexity of updating $\bt$ from Step 8 to Step 11 is $\mathcal{O}\left(M N L_{\rm_t} \gamma_{\rm_t}^1+N^{2.5} \ln (1 / \beta) \gamma_{\rm_t}^2\right)$. Hence, assuming the maximum number of outer iterations is $\gamma$, the total computational complexity is $\mathcal{O}((N^2 L_{\rm_t}+M N L_{\rm_r} \gamma_{\rm_r}^1+M^{2.5} \ln (1 / \beta) \gamma_{\rm_r}^2+M N L_{\rm_t} \gamma_{\rm_t}^1+N^{2.5} \ln (1 / \beta)\gamma_{\rm_t}^2)\gamma)$.

\section{Simulations Results}


In this section, we provide numerical analysis to show the performance of the proposed statistical CSI-aided
algorithm. We consider a MIMO system, including a transmitter and a receiver each with 4 FAs. The number of transmit and receive paths is set as $L_{\rm_t}=L_{\rm_r}=3$. We assume that the elevation and azimuth AoDs/AoAs $\{\theta_{\rm t}^p\}_{p=1}^{L_{\rm_t}}$, $\{\phi_{\rm t}^p\}_{p=1}^{L_{\rm_t}}$, $\{\theta_{\rm r}^q\}_{q=1}^{L_{\rm_r}}$, $\{\phi_{\rm r}^q\}_{q=1}^{L_{\rm_r}}$ are all independent and identically distributed variables randomly distributed in $[0,\pi]$. The variance of $\boldsymbol\Sigma_{i,j}$ is set as  $\alpha^2=1/L_{\rm_r}$. The signal wavelength is set as $\lambda=1.5$ m. The minimum required distance between transmit/receive FAs is set as $D=\lambda/2$. The average power of noise is set as $\sigma^2=15$ dBm. The signal-to-noise (SNR) is defined as $P_{\text{max}}/\sigma^2$. 

\begin{figure}[htbp]
	\centering
	\includegraphics[scale=0.5]{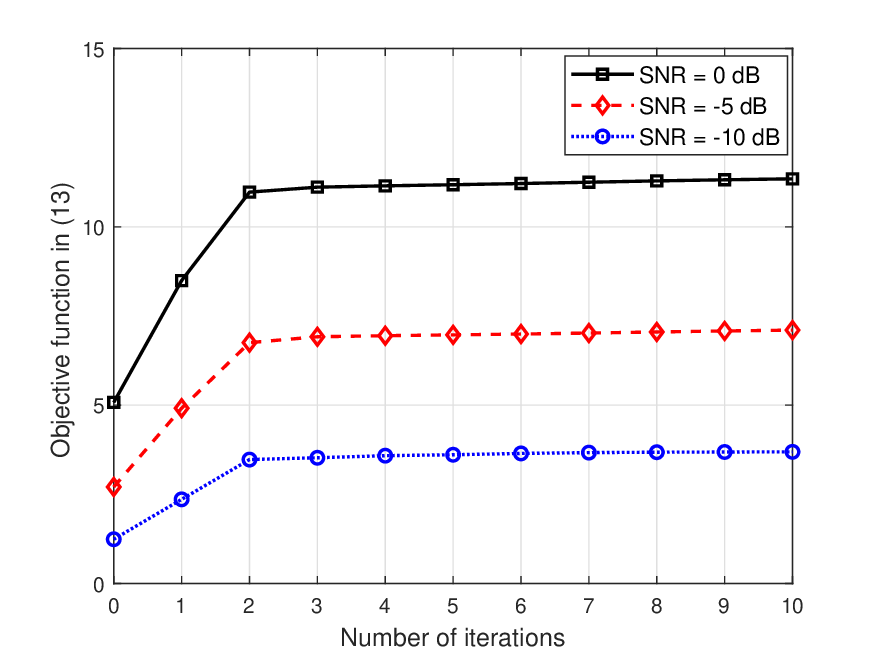}
	\captionsetup{font=footnotesize,singlelinecheck=off}
	\caption{Convergence of \textbf{Algorithm 1} under different values of $P_{\text{max}}$.}\label{fig}
\end{figure}
\begin{figure}[htbp]
	\centering
	\includegraphics[scale=0.5]{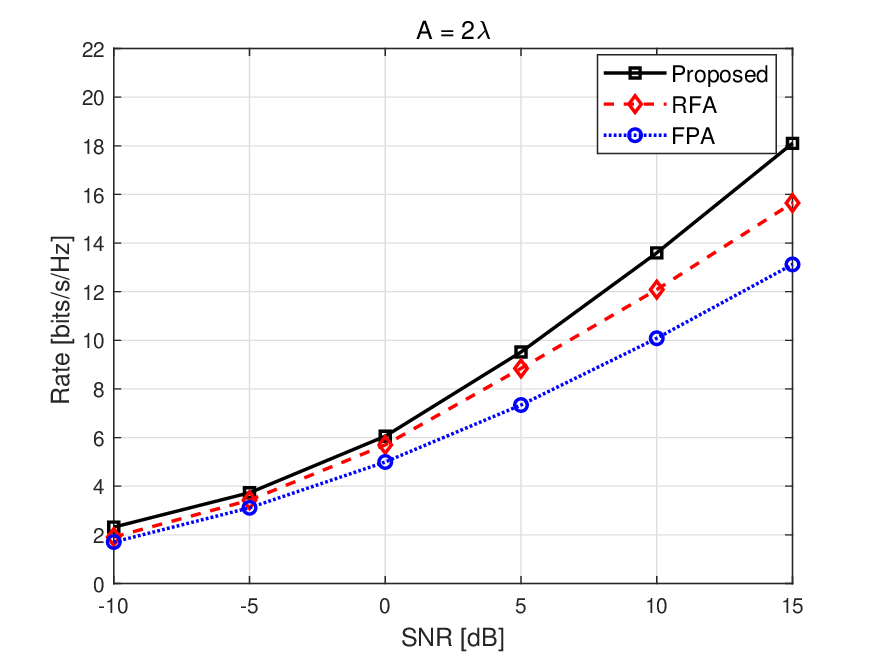}
	\captionsetup{font=footnotesize,singlelinecheck=off}
	\caption{Rate with respect to SNR.}\label{fig1}
\end{figure}
\begin{figure}[htbp]
	\centering
	\includegraphics[scale=0.5]{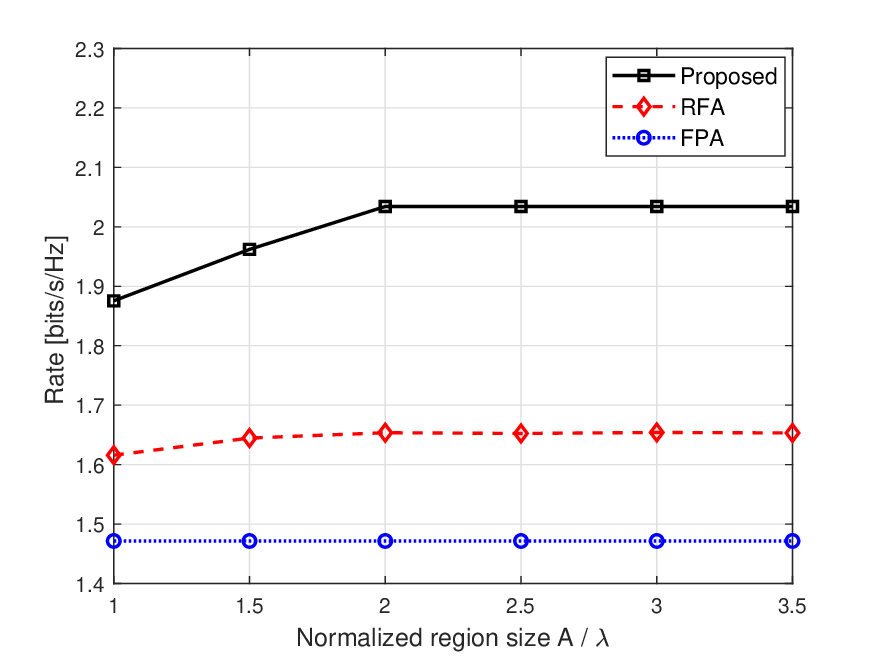}
	\captionsetup{font=footnotesize,singlelinecheck=off}
	\caption{Rate versus the size of the normalized region $A/\lambda$.}\label{fig2}
\end{figure}

We compare the proposed design with the following baselines:

\textbf{FPA}: The antennas at the transmitter and receiver are uniformly linear arrays and the inter-element antenna
spacing is $\lambda/2$.

\textbf{RFA}: The antennas at the transmitter are uniformly linear arrays, and the inter-element antenna spacing is $\lambda/2$. The antennas at the receiver are FAs, which can move freely in the given region.

The convergence performance of \textbf{Algorithm 1} under different values of $P_{\text{max}}$ is presented in Fig. \ref{fig}. It is shown that \textbf{Algorithm 1} can converge after a few iterations. 



Fig. \ref{fig1} compares the achievable rates of the FA-assisted MIMO system and baselines with respect to SNR. We can observe that for different SNR, the rate performance of the proposed design always outperforms the RFA and FPA designs, and the performance gain rises with the increase of SNR. Specifically, when SNR = 15 dB, the proposed FA design obtains 38.04$\%$ and 15.72$\%$ performance gains compared with the FPA and RFA designs.

Fig. \ref{fig2} illustrates the achievable rates of the FA-assisted MIMO system and baselines with respect to region size (normalized by wavelength) $A/\lambda$. It is observed that all FA-based approaches (either RFA or our proposed scheme) benefit from a larger antenna region. When the normalized region $A/\lambda$ size reaches 2, we can observe that curves corresponding to FA and RFA converge, which means that we can use a limited region of transmit/receive FAs to achieve the maximum achievable rate. Besides, it is clearly shown that compared with the RFA and FPA designs, the proposed FA design dramatically improves the rate performance of the MIMO system. Specifically, when $A = 3.5\lambda$, the proposed FA design obtains 34.72$\%$ and 23.68$\%$ performance gains compared with the FPA and RFA designs, respectively.

\section{Conclusion}\label{sec_conclusion} 
In this letter, we investigated rate maximization for a FA-assisted MIMO system with statistical CSI under the minimum inter-antenna distance and power constraints. To simplify the objective function, Jensen's inequality was used to derive an analytical upper bound for the ergodic achievable rate. Then an alternating optimization approach was utilized to tackle the rate maximization problem. We investigated the relationship between the achievable performance and system parameters, such as the normalized region size and SNR. Numerical results demonstrated the performance gains of the FA-assisted system over the conventional FPA counterpart. Moreover, the results
revealed that a limited region of transmit/receive FAs can be used to achieve the maximum achievable rate.

\bibliographystyle{IEEEtran}

\begin{thebibliography}{10}
\providecommand{\url}[1]{#1}
\csname url@samestyle\endcsname
\providecommand{\newblock}{\relax}
\providecommand{\bibinfo}[2]{#2}
\providecommand{\BIBentrySTDinterwordspacing}{\spaceskip=0pt\relax}
\providecommand{\BIBentryALTinterwordstretchfactor}{4}
\providecommand{\BIBentryALTinterwordspacing}{\spaceskip=\fontdimen2\font plus
\BIBentryALTinterwordstretchfactor\fontdimen3\font minus
  \fontdimen4\font\relax}
\providecommand{\BIBforeignlanguage}[2]{{%
\expandafter\ifx\csname l@#1\endcsname\relax
\typeout{** WARNING: IEEEtran.bst: No hyphenation pattern has been}%
\typeout{** loaded for the language `#1'. Using the pattern for}%
\typeout{** the default language instead.}%
\else
\language=\csname l@#1\endcsname
\fi
#2}}
\providecommand{\BIBdecl}{\relax}
\BIBdecl

\bibitem{wang2023road}
C.-X. Wang {\em et al.}, ``On the road to 6G: Visions, requirements, key technologies and testbeds,'' \emph{IEEE Commun. Surveys Tuts.}, vol. 25, no. 2, pp. 905--974, Feb. 2023.

\bibitem{wong2020fluid2}
K.-K. Wong, A.~Shojaeifard, K.-F. Tong, and Y.~Zhang, ``Fluid antenna systems,'' \emph{IEEE Trans. Wireless Commun.}, vol.~20, no.~3, pp. 1950--1962, Mar. 2021.

\bibitem{shojaeifard2022mimo}
A.~Shojaeifard {\em et al.}, ``{MIMO} evolution beyond {5G} through reconfigurable intelligent surfaces and fluid antenna systems,'' \emph{Proc. IEEE}, vol. 110, no.~9, pp. 1244--1265, Sep. 2022.

\bibitem{wong2023fluid00}
K.-K. Wong, W.~K. New, X.~Hao, K.-F. Tong, and C.-B. Chae, ``Fluid antenna system--{P}art {I}: Preliminaries,'' \emph{IEEE Commun. Lett.}, vol. 27, no. 8, pp. 1919--1923, Aug. 2023.

\bibitem{new2023information}
W.~K. New, K.-K. Wong, X.~Hao, K.-F. Tong, and C.-B. Chae, ``An information-theoretic characterization of {MIMO}-{FAS}: Optimization, diversity-multiplexing tradeoff and $ q $-outage capacity,'' accepted in {\em IEEE Trans. Wireless Commun.}, [Online] \url{arXiv:2303.02269}, 2023.

\bibitem{wong2020performance}
K.-K. Wong, A.~Shojaeifard, K.-F. Tong, and Y.~Zhang, ``Performance limits of fluid antenna systems,'' \emph{IEEE Commun. Lett.}, vol.~24, no.~11, pp. 2469--2472, Nov. 2020.

\bibitem{mukherjee2022level}
P.~Mukherjee, C.~Psomas, and I.~Krikidis, ``On the level crossing rate of fluid antenna systems,'' in \emph{Proc. IEEE 23rd Int. Workshop Signal Process.
  Adv. Wireless Commun. (SPAWC)}, Jul. 2022, pp. 1--5.

\bibitem{ma2022mimo}
W.~Ma, L.~Zhu, and R.~Zhang, ``{MIMO} capacity characterization for movable antenna systems,'' \emph{IEEE Trans. Wireless Commun.}, pp. 1--1, Sep. 2023.

\bibitem{zhu2022modeling}
L.~Zhu, W.~Ma, and R.~Zhang, ``Modeling and performance analysis for movable antenna enabled wireless communications,'' [Online] \url{arXiv:2210.05325}, 2022.

\bibitem{skouroumounis2022fluid}
C.~Skouroumounis and I.~Krikidis, ``Fluid antenna with linear {MMSE} channel estimation for large-scale cellular networks,'' \emph{IEEE Trans. Commun.}, vol.~71, no.~2, pp. 1112--1125, Feb. 2022.

\bibitem{ma2023compressed}
W.~Ma, L.~Zhu, and R.~Zhang, ``Compressed sensing based channel estimation for movable antenna communications,'' \emph{IEEE Commun. Lett.}, vol.~27, no.~10, pp. 2747--2751, Oct. 2023.

\bibitem{you2020energy}
L.~You {\em et al.}, ``Energy efficiency optimization for downlink massive {MIMO} with statistical {CSIT},'' \emph{IEEE Trans. Wireless Commun.}, vol.~19, no.~4, pp. 2684--2698, Apr. 2020.

\bibitem{9238778}
Y.~Huang, L.~You, J.~Xiong, W.~Wang, and X.~Gao, ``Max-min energy-efficient multi-cell massive {MIMO} transmission exploiting statistical {CSI},'' in \emph{Proc. IEEE/CIC Int. Conf. Commun. China (ICCC)}, Chongqing, China, Nov. 2020, pp. 330--335.

\bibitem{you2020spectral}
L.~You, J.~Xiong, A.~Zappone, W.~Wang, and X.~Gao, ``Spectral efficiency and energy efficiency tradeoff in massive {MIMO} downlink transmission with statistical {CSIT},'' \emph{IEEE Trans. Signal Process.}, vol.~68, pp. 2645--2659, Apr. 2020.

\bibitem{8926431}
A.~Arora, C.~G. Tsinos, B.~S. M.~R. Rao, S.~Chatzinotas, and B.~Ottersten, ``Hybrid transceivers design for large-scale antenna arrays using majorization-minimization algorithms,'' \emph{IEEE Trans. Signal Process.}, vol.~68, pp. 701--714, Dec. 2020.

\bibitem{9505311}
Y.~Chen, Y.~Wang, and L.~Jiao, ``Robust transmission for reconfigurable intelligent surface aided millimeter wave vehicular communications with statistical {CSI},'' \emph{IEEE Trans. Wireless Commun.}, vol.~21, no.~2, pp. 928--944, Aug. 2022.

\bibitem{8409338}
Q.~Qin, L.~Gui, P.~Cheng, and B.~Gong, ``Time-varying channel estimation for millimeter wave multiuser {MIMO} systems,'' \emph{IEEE Trans. Veh. Technol.}, vol.~67, no.~10, pp. 9435--9448, Jul. 2018.

\bibitem{mckay2006random}
M.~R. McKay, ``Random matrix theory analysis of multiple antenna communication systems,'' Ph.D. dissertation, School Elect. Inf. Eng., Telecommun. Lab., Faculty Eng. Inf. Technol., Univ. Sydney, 2006.

\bibitem{bertsekas1997nonlinear}
D.~P. Bertsekas, ``Nonlinear programming,'' \emph{J. Oper. Res. Soc.}, vol.~48, no.~3, pp. 334--334, 1997.

\bibitem{yang2001matrix}
X.~M. Yang, X.~Q. Yang, and K.~L. Teo, ``A matrix trace inequality,'' \emph{J. Math. Anal. Appl.}, vol. 263, no.~1, pp. 327--331, 2001.
\end{thebibliography}


\end{document}